\documentclass{PoS}

\usepackage{amssymb}
\usepackage{amsmath}

\definecolor{gray}{rgb}{0.6,0.6,0.6}
\definecolor{yellowgreen}{rgb}{0.55,0.8,0.2}
\definecolor{dimgray}{rgb}{0.4,0.4,0.4}
\definecolor{steelblue}{rgb}{0.27,0.51,0.7}
\definecolor{orange}{rgb}{1,0.65,0}
\definecolor{cornsilk}{rgb}{1,0.97,0.86}

\newcommand{\tanb}{\tan\beta}
\newcommand{\gev}{~\mathrm{GeV}}
\newcommand{\tev}{~\mathrm{TeV}}
\newcommand{\eg}{\textit{e.g.}}
\newcommand{\ie}{\textit{i.e.}}
\newcommand{\br}{\mathrm{BR}}
\newcommand{\HB}{\texttt{HiggsBounds}}
\newcommand{\HS}{\texttt{HiggsSignals}}

\title{Recent Developments in \HB~and a Preview of \HS}

\ShortTitle{Recent Developments in \HB~and a Preview of \HS}

\author{Philip Bechtle$^a$, Oliver Brein$^b$, Sven Heinemeyer$^c$, Oscar St\r{a}l$^d$, \speaker{Tim Stefaniak}$^a$, Georg Weiglein$^e$, Karina Williams$^a$\\
\llap{$^a$}  Physikalisches Institut der Universit\"at Bonn, Nu{\ss}allee 12, D-53115 Bonn, Germany\\
\llap{$^b$} Grosskarlbacher Stra{\ss}e 10, D-67256 Weisenheim am Sand, Germany\\
\llap{$^c$} Instituto de F\'isica de Cantabria (CSIC-UC), Santander, Spain\\
\llap{$^d$} The Oscar Klein Centre, Department of Physics, Stockholm University,\\SE-106 91 Stockholm, Sweden\\
\llap{$^e$} Deutsches Elektronen-Synchrotron DESY, Notkestra{\ss}e 85, D-22607 Hamburg, Germany\\
E-mail: \email{bechtle@physik.uni-bonn.de}, \email{oliver@brein.de}, \email{Sven.Heinemeyer@cern.ch}, \email{oscar.stal@fysik.su.se}, \email{tim@th.physik.uni-bonn.de}, \email{Georg.Weiglein@desy.de}, \email{williams@th.physik.uni-bonn.de} }

\abstract{
We report on recent developments in the public computer code \HB, which confronts arbitrary Higgs sector predictions with $95\%~C.L.$ exclusion limits from Higgs searches at the LEP, Tevatron and LHC experiments. We discuss in detail the performance of the Standard Model (SM) likeness test as implemented in the latest version \HB\texttt{-3.8.0}, whose outcome decides whether a search for a SM Higgs boson can be applied to a model beyond the SM. Furthermore, we give a preview of features in the upcoming version \HB\texttt{-4.0.0} and the new program \HS, which performs a $\chi^2$ test of Higgs sector predictions against the signal rate and mass measurements from Higgs boson analyses at the Tevatron and LHC. This is illustrated with an example where the heavier $CP$-even Higgs boson of the Minimal Supersymmetric Standard Model (MSSM) is considered as an explanation of the LHC Higgs signal at $\simeq 126\gev$.
}

\FullConference{Prospects for Charged Higgs Discovery at Colliders\\
                 October 8-11, 2012\\
                 Uppsala University Sweden}

\dedicated{BONN-TH-2013-01, DESY 13-004}

\begin{document}

%--------------------------------------------------------------------------------------------------------------
\section{Introduction to \HB}
\label{Sect:Intro}
%--------------------------------------------------------------------------------------------------------------
The search for the Higgs boson at collider experiments is essential for unravelling the mechanism of electroweak symmetry breaking. These searches have been carried out over the last two decades at the LEP and Tevatron experiments, and for the last two years also at the LHC experiments. Negative search results\footnote{We discuss the recent discovery in the Higgs searches at the LHC in Sect.~\ref{Sect:newdevs}.} are used to derive $95\%~C.L.$ limits on the Higgs signal rate. These results are provided either as fairly \textit{model-independent} limits on the cross section of a single signal topology (\eg~the LEP process $e^+e^- \to h_i Z \to b\bar{b}Z$ or the charged Higgs process in top quark production, $t\to H^+ b \to \tau \nu_\tau b$) or as a limit on combined cross sections for a specific model, in particular the Standard Model (SM). In the latter case, the discovery potential is maximized by the combination of several production and/or decay modes, at the cost of a more \textit{model-dependent} cross section limit.

\begin{figure}[b]
\centering
\includegraphics[width=0.5\textwidth]{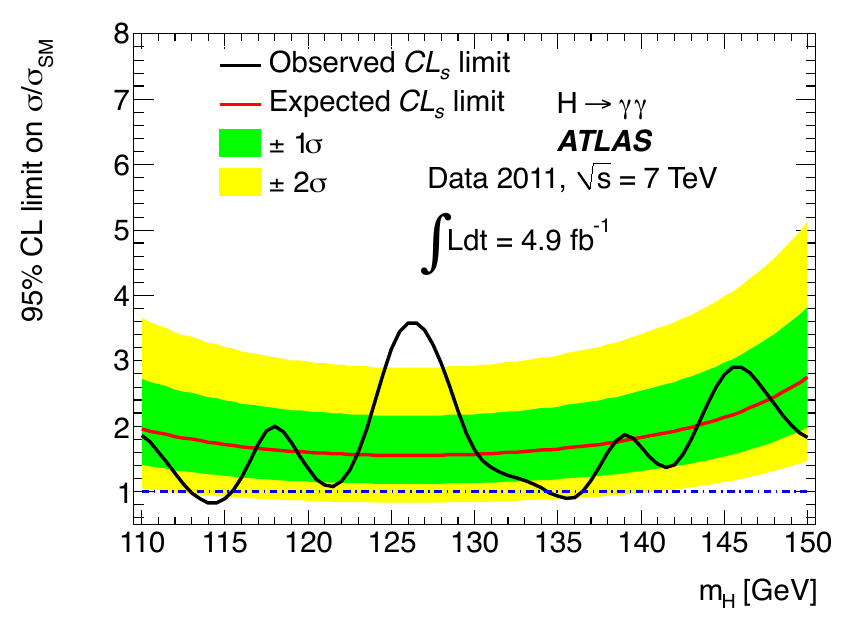}
\caption{$95\%~C.L.$ exclusion limit from the ATLAS Standard Model $H\to\gamma\gamma$ search~\cite{ATLAS:2012ad}. The limit is given on the signal strength modifier, $\mu = \sigma/\sigma_\mathrm{SM}$, which universally scales the Standard Model Higgs boson signal rates of the considered production and decay modes.}
\label{Fig:ATLASexcl}
\end{figure}

The limit on the Higgs boson signal rate is derived by rejecting the Higgs signal plus background hypothesis. An example of a model-dependent exclusion limit is given in Fig.~\ref{Fig:ATLASexcl} for the ATLAS search for a SM Higgs boson decaying to two photons~\cite{ATLAS:2012ad}. The figure presents both the observed limit and the expected limit on a universal scale factor $\mu=\sigma/\sigma_\mathrm{SM}$ of the combined SM Higgs boson signal rate as a function of the assumed Higgs mass $m_H$. The expected limit is derived from Monte Carlo simulation where the data is assumed to be identical to the background expectation. In this example, the considered Higgs production modes comprise of the gluon-gluon fusion (ggf) and vector boson fusion (VBF) process as well as the associated Higgs production with a vector boson ($HW$, $HZ$) and with a top quark pair ($Ht\bar{t}$). Without the knowledge of the individual efficiencies of these signal topologies, the limit applies only to models where the five production modes are approximately in the same relations as in the SM. This is discussed in detail in Section~\ref{Sect:SMtest}.

The public computer code \HB~\cite{Bechtle:2008jh, Bechtle:2011sb, HB3manual} confronts arbitrary Higgs sector predictions with the most recent $95\%~C.L.$ exclusion limits from neutral and charged Higgs searches from the LEP, Tevatron and LHC experiments. In order to preserve the $95\%~C.L.$ limit interpretation, we follow a well-defined statistical procedure: First, the most sensitive analysis (\ie~the analysis with the best \textit{expected} exclusion limit for the particular parameter point of the investigated model) is determined out of all available analyses. Then, the parameter point is tested against the \textit{observed} limit of this particular analysis \textit{only}.

As input the code requires the number of neutral and charged Higgs bosons $h_i$, their masses $m_{h_i}$, total widths $\Gamma_\mathrm{tot}(h_i)$, decay branching ratios $\br(h_i \to \dots)$, (normalized) production cross sections as well as the top quark branching ratios. The latter are needed for limits from light charged Higgs searches in top quark decays. Various input approximations are available, \eg~the production cross sections and branching ratios to SM particles can be derived from effective Higgs couplings given by the user. \HB~can be called from the command-line, via \texttt{Fortran90} subroutines %\footnote{A \texttt{Fortran77} code was maintained until version \HB\texttt{-3.7.0}.}
or from a web-interface. It supports\footnote{The required \HB~effective coupling input has to be provided via two extra SLHA blocks, see Ref.~\cite{HB3manual}.} the SUSY Les Houches Accord (SLHA) for the Minimal Supersymmetric Standard Model (MSSM) and the next-to-MSSM (NMSSM). The narrow-width approximation is assumed to be applicable. The \HB~result contains information about whether the model is excluded at $95\%~C.L.$ and which analysis was applied to which Higgs boson.

The current version is \HB\texttt{-3.8.0} and was released in May 2012. The included LHC results are from the $7\tev$ run only. The code can be obtained from

\begin{center}
\texttt{http://higgsbounds.hepforge.org}
\end{center}

A new version \HB\texttt{-4.0.0} including the latest $8\tev$ LHC results will be released soon. More details will be given in Section~\ref{Sect:newdevs}.

%--------------------------------------------------------------------------------------------------------------
\section{Performance of the Standard Model likeness test}
\label{Sect:SMtest}
%--------------------------------------------------------------------------------------------------------------

If a Higgs search is carried out under specific assumptions (\eg~on the $CP$-properties of the Higgs boson or top quark branching ratios) \HB~considers these analyses only if the investigated model fulfills these assumptions. In particular, as mentioned in Section~\ref{Sect:Intro}, if the Higgs search combines several signal topologies under the assumption of the SM, \HB~tests whether the model is sufficiently SM-like in these signal topologies. This \textit{SM-likeness test} is necessary since the information on efficiencies of the various signal topologies is rarely made publicly available.

For each signal topology $i$, comprised of the production mode $P(h)$ and decay to the final state $F$, we define an individual signal strength modifier $c_i$ and SM weight $\omega_i$,
\begin{eqnarray}
c_i = \frac{[\sigma_\mathrm{model}(P(h)) \br_\mathrm{model}(h\to F)]_i}{[\sigma_\mathrm{SM}(P(H)) \br_\mathrm{SM}(H\to F)]_i},
\qquad
\omega_i = \frac{[\sigma_\mathrm{SM}(P(H)) \br_\mathrm{SM}(H\to F)]_i}{\sum_j [\sigma_\mathrm{SM}(P(H)) \br_\mathrm{SM}(H\to F)]_j}.
\end{eqnarray}
Here, the sum runs over all considered signal topologies $j$ of the analysis. Thus, the SM weight $\omega_i$ describes the relative contribution of the signal topology $i$ to the total signal rate in the SM. 

\begin{figure}
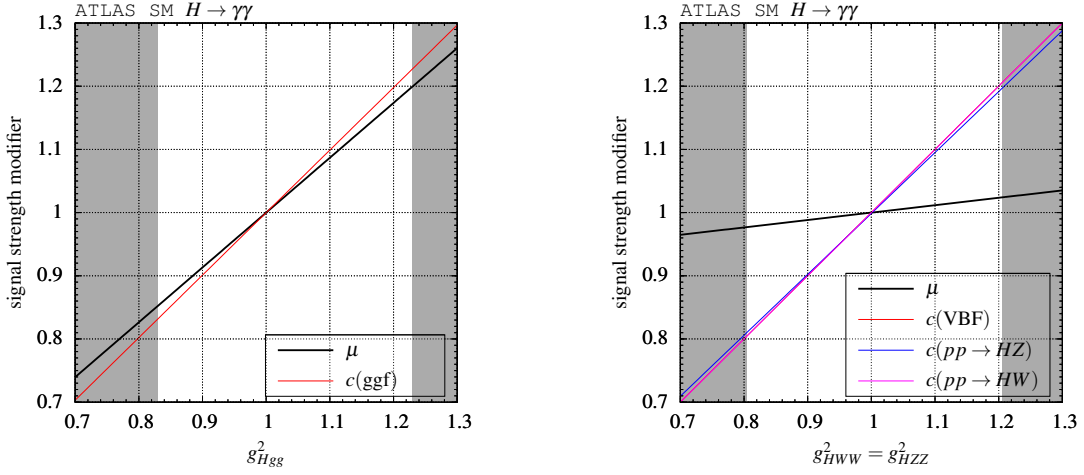

\hspace{-1.0cm}\scalebox{0.7}{\input{1414_mh125_g2hgg_weights}}
\hspace{-1.0cm}\scalebox{0.7}{\input{1414_mh125_g2hWW_ZZ_weights}}
\caption{Performance of the SM likeness test: We consider the ATLAS $H\to\gamma\gamma$ search~\cite{ATLAS:2012ad} and modify the SM normalized squared effective Higgs couplings $g^2_{Hgg}$ (\textit{left}) and $g_{HVV}^2$ ($V=W,Z$) (\textit{right}) for a Higgs boson with mass $m=125\gev$. We show the dependence of the total signal strength modifier $\mu$ and the relevant individual signal strength modifiers. The gray regions indicate the parameter space where the SM likeness test fails.}
\label{Fig:SMtest}
\end{figure}

Neglecting the effects from different efficiencies among the signal topologies, the total signal strength modifier $\mu$ is approximated by $\mu = \sum_i \omega_i c_i$. However, this approximation is only valid if the signal topologies in the model contribute in similar proportions to the total signal rate as in the SM, or in other words, as long as the individual signal strength modifiers $c_i$ are approximately the same as the total signal strength modifier $\mu$. Therefore, in order to pass the SM likeness test, we require the model to fulfill the following criterion:
\begin{eqnarray}
\Delta \equiv \max_{i}~\omega_i \left|\frac{\delta c_i}{\mu} \right| < \epsilon, \qquad \mbox{with} \qquad \delta c_i = c_i - \mu\qquad \mbox{and} \qquad \epsilon = 2\%,
\label{Eq:SMtest}
\end{eqnarray}
\ie~the maximal weighted deviation of the individual signal strength modifier from the total signal strength modifier is required to be less than $2\%$. We consider $\epsilon = 2\%$ as a conservative choice, considering that the uncertainties on the rate predictions are generally larger.

An essential part in the calculation of $\Delta$ is the inclusion of the SM weight $\omega_i$, which we want to illustrate with an example. We consider again the ATLAS $H\to \gamma\gamma$ search~\cite{ATLAS:2012ad} and test a model with a Higgs boson with a mass $m=125\gev$. We depart from the SM by modifying either the (SM normalized) squared effective Higgs coupling to gluons, $g_{Hgg}^2$, or to vector bosons, $g_{HVV}^2$ ($V=W,Z$). All other effective Higgs couplings, in particular the $H\gamma\gamma$ coupling, are set to their SM values. At $m=125\gev$, the SM weights for the LHC at $\sqrt{s}=7\tev$ are
\begin{eqnarray}
\omega \approx (87.7\%,~6.8\%,~3.2\%,~1.8\%,~0.5\%) \quad \mbox{for}\quad (\mathrm{ggf},~\mathrm{VBF},~HW,~HZ,~Ht\bar{t}).
\end{eqnarray}
In Fig.~\ref{Fig:SMtest} we show the signal strength modifier $\mu$ and the $c_i$ for the signal topologies influenced by the modified effective Higgs couplings. Varying $g^2_{Hgg}$ influences only the gluon-gluon fusion cross section, however, due to its large SM weight, $\omega_\mathrm{ggf} \approx 87.7\%$, the total signal strength modifier $\mu$ follows closely $c (\mathrm{ggf})$. The SM likeness test failure at $g^2_{Hgg} = 0.835$ and $1.225$ is eventually caused by the ggf signal topology, although the deviation $\delta c_i$ for the remaining signal topologies VBF, $HW$, $HZ$ and $Ht\bar{t}$ is much larger here. However, the SM weights of these channels are much smaller, thus allowing for a larger deviation. The same effects can be seen when varying $g_{HVV}^2$ ($V=W,Z$). Now, the $c_i$ of the VBF, $HW$, $HZ$ signal topologies are affected by the modified effective coupling, however, the total signal strength modifier $\mu$ is only slightly influenced due the small weight of these channels. Again, the deviation between $\mu$ and $c(\mathrm{ggf})$ is eventually causing the SM likeness test to fail. Thus, due to the inclusion of the SM weights in Eq.~(\ref{Eq:SMtest}) subdominant signal topologies are allowed to deviate further from $\mu$.

The SM weights were introduced with the latest version \HB\texttt{-3.8.0}. This led to a wider applicability of SM Higgs search results to arbitrary Higgs sectors and thus to a significant improvement of the performance of \HB. 

%--------------------------------------------------------------------------------------------------------------
\section{Prospects and new developments}
\label{Sect:newdevs}
%--------------------------------------------------------------------------------------------------------------

In July 2012, the long-lasting effort of the Higgs searchers was rewarded with a discovery in the Standard Model (SM) Higgs searches for $H\to \gamma\gamma$ and $H\to ZZ^{(*)} \to 4\ell$ at a mass value $m\approx 126\gev$ in both LHC experiments ATLAS~\cite{ATLAS:2012gk} and CMS~\cite{CMS:2012gu}. The major tasks after the discovery are the determination of the couplings, $CP$ and spin properties of the new Higgs-like state. Any deviations from the SM expectation will hint towards new physics beyond the Standard Model and must be investigated thoroughly.

The \HB~team is currently developing the program \HS, aimed to perform a $\chi^2$ test of Higgs sector predictions against the signal rate and mass measurements from Higgs collider searches. This information can for instance be used in global fits of models beyond the SM or in a generic Higgs coupling determination. In the $\chi^2$ calculation the systematic uncertainties of production cross sections, decay rates, luminosities and Higgs mass predictions are taken into account with the full information on their correlations. Furthermore, the code automatically considers superpositions of the signal rates of Higgs bosons, if their mass difference cannot be resolved by the experimental analysis.

\begin{figure}
\centering
\vspace{-0.6cm}
\includegraphics[width=0.68\textwidth]{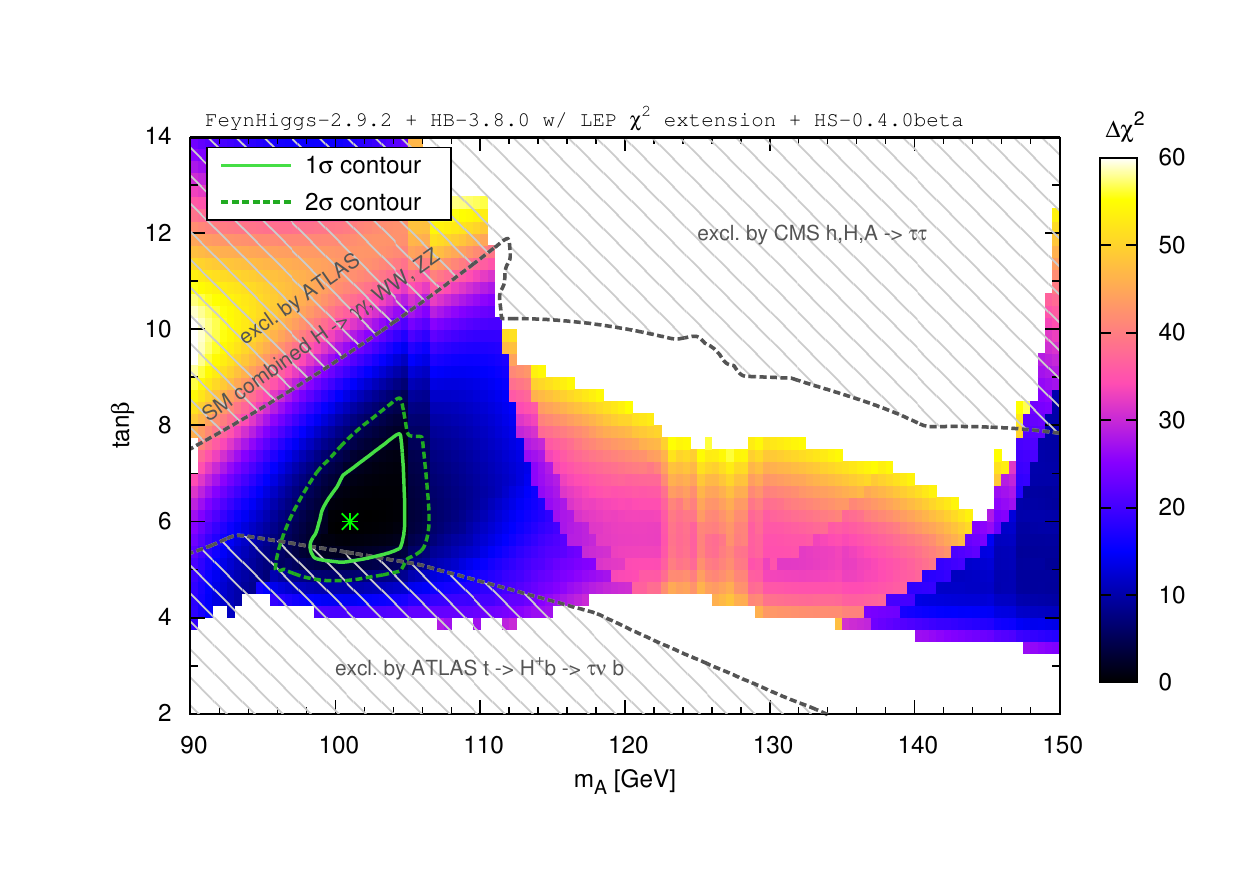}
\vspace{-0.2cm}
\caption{$\chi^2$ distribution for the heavy $CP$-even Higgs boson interpretation of the LHC Higgs signal in the ($m_A,~\tanb$) plane. The gray striped regions are excluded at $95\%~C.L.$ by \HB. The best-fit point is indicated by a green asterisk.}
\label{Fig:heavyH}
\end{figure}

To illustrate some features of the upcoming programs \HB\texttt{-4.0.0} and \texttt{Higgs\-Signals-1.0.0} we discuss the interesting possibility of interpreting the LHC Higgs signal as the heavier $CP$-even Higgs boson $H$ of the MSSM (see also Ref.~\cite{Heinemeyer:2011aa,Drees:2012fb,Bechtle:2012jw}~for a detailed analysis). We consider the ($m_A,~\tanb$) plane for the fixed choice of MSSM parameters
\begin{eqnarray*}
M_{SUSY} = 1\tev,~|X_t|=2.4\tev,~\mu=1\tev,~M_1=100\gev,~M_2 = 200\gev,~M_3=800\gev.
\end{eqnarray*}
A definition of these parameters can \eg~be found in Ref.~\cite{Heinemeyer:2011aa}. In Fig.~\ref{Fig:heavyH} we show the total $\Delta\chi^2$ likelihood map for this scenario. It is composed of the $\chi^2$ calculated by \HS~using the signal rate and mass measurements presented at the ICHEP2012 conference by ATLAS~\cite{ATLAS:2012gk}, CMS~\cite{CMS:2012gu}, D\O~and CDF~\cite{TCB:2012zzl}, and a $\chi^2$ value from the LEP exclusion. This LEP $\chi^2$ information will be included in \HB\texttt{-4.0.0}. Furthermore we show the $95\%~C.L.$ excluded regions obtained with \HB\texttt{-3.8.0}. While the ATLAS SM combined analysis of the channels $H \to \gamma\gamma,~WW,~ZZ$~\cite{ATLAS:2012ae} and the CMS $h,H,A\to \tau\tau$ search~\cite{Chatrchyan:2012vp} constrain\footnote{Note that the newer $95\%~C.L.$ exclusion limits from the LHC $8\tev$ run are even more constraining than the exclusion limits applied here.} the parameter regions with $\tanb \gtrsim 8$, the ATLAS charged Higgs search~\cite{Aad:2012tj} excludes regions at low $m_A$ and $\tanb$.

We find the best-fit point at $(m_A,~\tanb) = (101.0\gev,~6.0)$ with a good fit quality $\chi^2/\mathrm{ndf} = 29.3/32$. It exhibits a charged Higgs boson $H^\pm$ and a light $CP$-even Higgs boson $h$ with masses $m_{H^\pm}=126\gev$ and $m_h=92.3\gev$, respectively. The light Higgs boson $h$ has a reduced coupling to $Z$ bosons, hence it escapes the LEP exclusion limits. The rate for the charged Higgs process in top decays, $t \overset{1\%}{\rightarrow} H^+ b \overset{98\%}{\rightarrow} \tau^+ \nu_\tau b$, is fairly close the ATLAS exclusion limit~\cite{Aad:2012tj}. Thus, with more data charged Higgs boson searches will be able to probe this interpretation of the LHC Higgs signal.

\end{document}